# Factors affecting the Optimal Design of High-$T_c$ Superconductors – the Pseudogap and Critical Currents†

J. L. Tallon[a], G.V.M. Williams[a] and J.W. Loram[b]

[a]New Zealand Institute for Industrial Research, P.O. Box 31310, Lower Hutt, New Zealand.
[b]IRC in Superonductivity, Cambridge University, Cambridge CB3 0HE, England

**Abstract**

The impact of the normal-state pseudogap, present in all optimal and underdoped HTS cuprates, on critical currents and critical temperature is surveyed. With the opening of the pseudogap around a doping state of p≈0.19 the condensation energy and superfluid density are rapidly suppressed due to reduction in the normal-state spectral weight. Even by optimal doping (p≈0.16) these measures of the 'strength' of superconductivity are diminished by up to 40%. This results in a sharp reduction in critical currents and irreversibility field, respectively. The optimal doping state where these properties are maximised is therefore not at maximum $T_c$ but in the lightly overdoped region where the pseudogap energy falls to zero at p≈0.19. The presence of impurities and grain boundaries further heightens these effects.

## 1. Introduction

The optimal design of high-$T_c$ superconducting (HTS) cuprates, in particular the maximising of both critical current density, $J_c$, and critical temperature, $T_c$, may be considered in the context of a variety of structural, chemical, physical and materials science issues. These are governed by three central factors: (i) the generic doping dependence of the quasi-2D physical properties associated with the $CuO_2$ planes; (ii) the 3D anisotropy that describes the interlayer coupling between successive stacks of $CuO_2$ planes, and (iii) extrinsic effects associated with defects, impurities and grain boundaries. $J_c$ proves to be a strong function of all three: doping, coupling and extrinsic effects. The HTS cuprates exhibit a generic phase behaviour as a function of hole concentration, p, ranging from an antiferromagnetic (AF) insulator at zero doping to a metallic Fermi liquid at high doping with the appearance of superconductivity (SC) at intermediate doping levels. In this region $T_c$ follows a roughly parabolic dependence on hole concentration, p, approximated by [1]

$$T_c = T_{c,max} [1 - 82.6(p-0.16)^2] \qquad (1)$$

Elsewhere [2] we discussed the structural factors that govern the magnitude of $T_{c,max}$ in terms of correlations using copper and oxygen bond valence sum parameters, $V_{Cu}$ and $V_O$. These correlations indicate that structural changes which cause a displacement of positive charge from copper to oxygen (i.e. when the carriers have a stronger oxygen 2p character) will tend to increase $T_{c,max}$. Stategies to increase $T_{c,max}$, include increasing the apical-oxygen bond length, reducing the $CuO_2$ plane buckling angle, increasing the in-plane Cu-O bond length and moving the alkaline earth ion further away from the $CuO_2$ plane. Such displacements need to be effected while still maintaining optimal doping. Application of these ideas leads to the prediction that the highest $T_c$ to be seen in a cuprate will be found in the compound $HgRa_2Ca_2Cu_3O_8$ in which the radium ion will additionally stretch the CuO in plane bond while maintaining a high apical oxygen bond length.

---

†Based on an unpublished invited talk presented at the symposium on Processing and Critical Currents of HTS 2-4 Feb, 1998, Wagga Wagga, Australia .

Structural features also affect the anisotropy and hence the irreversibility field, $H_{irr}$. If one takes a fixed doping state, for example optimal doping, $p=0.16$, then $H_{irr}$ follows an exponential dependence upon the block-layer spacing, $d_b$, between the $CuO_2$ sheets [3]. A remarkable example of this dependence arises from metallising the block-layer. Proximity-induced superconductivity in the interlayer results in a strongly enhanced superfluid density with a consequent halving of the effective interlayer spacing. The exponential dependence on $d_b$ may result in a 20- to 50-fold enhancement in $H_{irr}$ [4]. Examples include full oxygenation of the $CuO_{1-\delta}$ chains in Y-123 and Y-247 and the substitution of 25% Re for Hg in Hg-1212 or Hg-1223. An attempt to employ this strategy by means of inserting a metallic $RuO_2$ layer between the $CuO_2$ layers in $RuSr_2GdCu_2O_8$ resulted in the unexpected discovery of coexisting superconductivity and ferromagnetism in this compound [5].

Strategies to enhance flux pinning include rapid phase conversion to induce structural defects. One such method employs the conversion of lightly La-doped Y-124 to Y-123 by exposure to low $PO_2$ for just a few minutes [6]. The extrinsic La atoms which substitute for Ba appear to pin a remnant of CuO double chains which terminate within the lattice, possibly adjacent to the substituent. These terminations provide pinning sites which enhance the critical current in Y-123 at 77K by an order of magnitude and by a factor of 50 at 87K.

These various approaches to material design reflect the fact that we understand the basic *structural, chemical* and *materials* principles of $T_c$ and $J_c$ enhancement in HTS cuprates. Some of this knowledge is new, while some is derived from our previous understanding of LTS. The challenge to solid-state chemists is whether this knowledge can be usefully implemented in the development of high-performance novel materials. The remainder of this paper deals with issues that emerge from recent developments in our understanding of the *physics* of the cuprates, primarily associated with the pseudogap which is present in all underdoped samples.

## 2. The pseudogap

Underdoped cuprates exhibit normal state (NS) correlations above $T_c$ which result in a depletion of the density of states (DOS), referred to as the *pseudogap*. This gap is pinned to the Fermi level, independent of doping [7], and therefore dominates the NS low-energy excitations. Observed first in NMR and inelastic neutron scattering and interpreted as a spin gap it was subsequently shown from heat capacity, then ARPES and tunneling, to be a gap in the spectrum of quasiparticles [8]. While many models have been proposed for understanding the pseudogap a widely-favoured scenario is that it arises from incoherent pairing fluctuations above $T_c$ [9], an outlook not shared by the present author. An alternative view is that the pseudogap is a competing correlation that takes away spectral weight otherwise available for superconductivity (SC). The distinction between these two scenarios is critical when it comes to practical SC properties such as $J_c$ for, if it is an independent correlation which competes with SC to the lowest temperatures, then its presence will be manifest in a strong reduction in pinning and $J_c$. If the pseudogap arises from phase fluctuations in a Cooper pairing state then at low enough temperature these fluctuations are no longer likely to be relevant and therefore the full rigidity of the SC wavefunction will be restored. Effects would then be confined to the doping-dependent reduction in carrier density and the associated (proportional) reduction in superfluid density. Evidence for the pseudogap being a competing correlation has been presented elsewhere [10,11]. Some of the key factors may be listed as follows. The pseudogap and SC energy scales ($E_g$ and $\Delta_o$) have different doping dependencies; the Knight shift and entropy scale to a universal curve above *or* below $T_c$ but not *both*; an isotope effect is observed in $T_c$ but not in the pseudogap; and, the magnitude of $E_g$ rises to over 1000K in heavily underdoped samples, too high for SC and clearly governed by the exchange energy J rather than a pairing strength. Moreover, the closure of the pseudogap appears to be directly associated with the disappearance of AF correlations and the occurrence of the NS metal/ insulator transition [12]. This would seem to be clear evidence that the pseudogap is a competing correlation, associated in some way with AF correlations. In the following we explore the consequences on $J_c$ of this competition.

## 3. Condensation energy

As a useful starting point we consider the p-dependent variation of the thermodynamic properties of the cuprates as determined by Loram and coworkers [7,10]. First we consider the variation of the jump, $\Delta\gamma_c$ in the specific heat coefficient $\gamma$ as a function of doping. Without resorting to specific models, a simple qualitative inspection of the data for Y-123, Y,Ca-123, La-214 and Bi-2212 reveals a drastic change in $\Delta\gamma_c$ at the critical doping state $p\approx 0.19$ [10]. With decreasing doping across the overdoped region $\Delta\gamma_c$ increases slowly, consistent with entropy balance and the increasing $T_c$. Then, beginning at $p=0.19$ in the lightly overdoped region, $\Delta\gamma_c$ falls sharply and even by optimal doping, $p=0.16$, it has decreased by 40% or so. This is due to the opening of the pseudogap and $\Delta\gamma_c$ goes on decreasing rapidly with further underdoping. Although pairing correlations above $T_c$ would also reduce $\Delta\gamma_c$ it is the suddenness of the onset in reduction that is difficult to understand within a phase-fluctuation scenario. Indeed, the argument for phase fluctuations was based on the observed low superfluid density in underdoped cuprates [9] but the sudden reduction in $\Delta\gamma_c$ at $p=0.19$ occurs at the very point in the entire phase diagram where the superfluid density is at its maximum value [13] (see Section 5). $\rho_s$ is found to fall away sharply on both sides of critical doping (see Fig. 4 later). The implication is that, but for the appearance of the pseudogap $\rho_s$ would continue to have increased with decreasing doping thus forstalling the onset of phase fluctuations if this were the origin of the pseudogap. The jump $\Delta\gamma_c$ is a measure of the pair density [14] but is not, in general, a T=0 property. More compelling is the behaviour of the condensation energy, $U_o$, defined by

$$U_o = \int_0^{T_c} [S_{NS} - S_{SC}]\, dT, \qquad (2)$$

where $S_{NS}$ is the extrapolated NS electronic entropy and $S_{SC}$ is the SC state electronic entropy. Fig. 1 shows values of $U_o(p)$ calculated from the electronic entropy for $Y_{0.8}Ca_{0.2}Ba_2Cu_3O_{7-\delta}$ [10]. The figure also shows the values of $E_g$ for the same samples obtained by scaling $S_{NS}/T$ to a universal function of $T/E_g$. This yields the doping dependence of the pseudogap without having to resort to a specific model. (A similar scaling below $T_c$ yields the p-dependence of $\Delta_o$). The figure shows that where $E_g$ falls to zero (at $p=0.19$) $U_o$ passes through a rather sharp maximum and we conclude that superconductivity is most robust at this critical doping point, not at optimal doping ($p\approx 0.16$) where $U_o$ has already halved. It should be remarked that it is only the behaviour for $p<0.19$ that is unconventional here. For a flat-band Fermi liquid one expects $U_o$ to scale as $\Delta_o^2$ and this appears to be well sustained. Fig. 2 shows that the p-dependence of $U_o/\Delta_o^2$ (open circles) remains constant on the overdoped side but, beginning at $p=0.19$ falls, sharply due to the opening of the pseudogap.

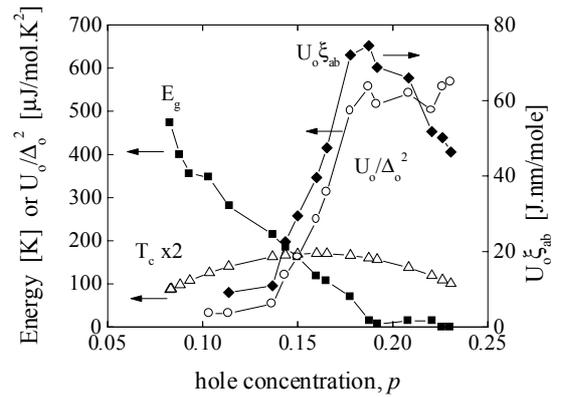

Fig. 2. $T_c$, $U_o/\Delta_o^2$, $U_o\times\xi_{ab}(0)$ and pseudogap energy $E_g$ determined from thermodynamic measurements on $Y_{0.8}Ca_{0.2}Ba_2Cu_3O_{7-\delta}$.

## 4. Flux pinning and critical current

The sharp peaking of $U_o$ at critical doping has important consequences for flux pinning and critical currents. A simple model of collective pin-ning leads to the conclusion that $J_c$ varies as the product of the gradient in pinning potential, $\nabla U_p$, times the coherence area, $\pi\xi^2$ [15]. The former may be approximated by $\nabla U_p \approx U_o/\xi$ with the result that $J_c$ varies as the product $U_o\xi$. The sharp peak in $U_o$ leads one therefore to expect a similar peak in the doping dependence of $J_c$ especially as the variation of $\xi(p)$ is not expected to be strong. This may be crudely estimated from the observed p-dependence of $\Delta_o$ using the relation $\xi = \hbar V_F/\pi\Delta_o$ where $V_F$ is the Fermi velocity. Assuming a 2D cylindrical Fermi surface with $V_F=(\hbar/m^*)(\pi n)^{0.5}$ and a resultant $\sqrt{p}$ variation, we obtain the p-dependence of $U_o\xi$ as shown in Fig.2. While the peak is not as sharp as that found for $U_o$ alone, $U_o\xi$ does indeed pass through an abrupt maximum at critical doping which we anticipate will be reflected in the $J_c$ variation with doping. This was investigated by magnetisation measurements.

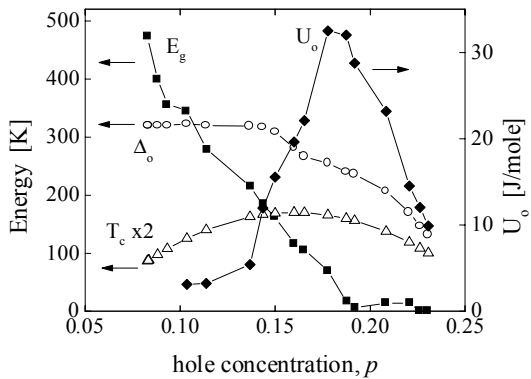

Fig. 1. $T_c$, condensation energy $U_o$ and pseudogap energy $E_g$ determined from thermodynamic measurements on $Y_{0.8}Ca_{0.2}Ba_2Cu_3O_{7-\delta}$.

A range of single-phase samples of $Y_{0.8}Ca_{0.2}Ba_2Cu_3O_{7-\delta}$ were annealed in various temperatures and oxygen partial pressures then quenched into liquid nitrogen to freeze in the oxygen content and doping state. These samples were extensively characterised using thermoelectric power, neutron diffraction, $^{89}Y$ NMR spectroscopy and were the subject of the thermodynamic studies summarised in Figs. 1 and 2. They were ground, milled and then c-axis aligned in epoxy in a magnetic field of 11.7 Tesla. In-plane magnetisation critical currents were then determined using the Bean model by measuring M-H loops at various temperatures with $H\|c$. $J_c(p)$ values [13] are plotted in Fig. 3 for H=0.2 Tesla at 10K and 20K. Each data set passes through a sharp maximum at critical doping reflecting the observed peak in $U_o\xi$. In view of the foregoing we may qualitatively understand this generic behaviour as follows. For $p>0.19$ $J_c(p)$ falls in conventional manner simply because $\Delta_o$ decreases ($J_c \sim U_o\xi \sim (U_o/\Delta_o^2) \times \Delta_o \sim \Delta_o$) but for $p<0.19$, where $\Delta_o$ remains more or less constant, $J_c(p)$ falls rapidly due to the opening of the pseudogap.

## 5. Superfluid density

The superfluid density, $\rho_s = n_s/m^* \propto \lambda_{ab}^{-2}$, is perhaps the most important property in defining the "strength" of superconductivity. Here $n_s$ is the density of carriers, $m^*$ is their effective mass and $\lambda_{ab}$ is the a-b plane penetration depth. The value of $\rho_s$ determines the rigidity of the condensate wavefunction, is central to the phenomenology of the irreversibility behaviour and is low enough in the cuprates to cause phase fluctuations [16] which reduce $T_c$ a small amount below its mean-field value. The superfluid density can be determined from the London penetration depth using ac suceptibility measurements or muon spin relaxation. In the later case $\lambda_{ab}^{-2} \propto \sigma_o$, the μSR depolarisation rate. As the pseudogap opens with progressive underdoping the magnitude of the superfluid density is sharply reduced and its temperature dependence is modified from its conventional d-wave behaviour [17]. Fig. 4 shows the p-dependence of $\sigma_o \propto \lambda_{ab}^{-2}$ determined from µSR measurements on $Y_{0.8}Ca_{0.2}Ba_2Cu_3O_{7-\delta}$ with different oxygenation levels. Data for 2% and 4% Zn substitution is also shown. Like $U_o$, $\lambda_{ab}^{-2}$ is observed to pass through a maximum at p=0.19 where the pseudogap first sets in, and by optimal doping has fallen some 40%. We will see that this doping dependence and in particular the effect of the pseudogap has a critical effect on irreversibility fields. Within the precursor pairing model [9] the reduction in superfluid density on the underdoped side due to the pseudogap and on the overdoped side, possibly due to pair-breaking [18], should have the effect of promoting phase fluctuations on both sides of optimal doping. This has, however, only been considered on the underdoped side [9].

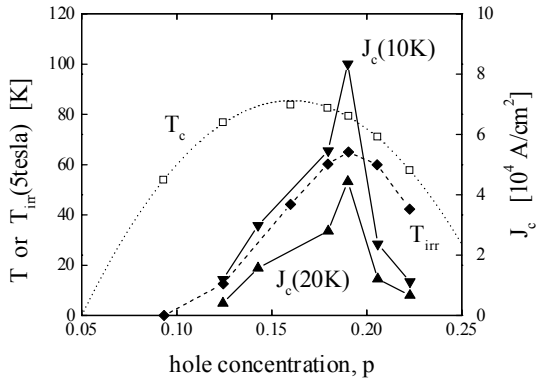

Fig. 3. The p-dependence of the magnetisation critical current density, $J_c(T=10K)$ and $J_c(T=20K)$ at 0.2Tesla for $Y_{0.8}Ca_{0.2}Ba_2Cu_3O_{7-\delta}$. The irreversibility temperature $T_{irr}(p)$ at 5 Tesla and $T_c$ are also shown.

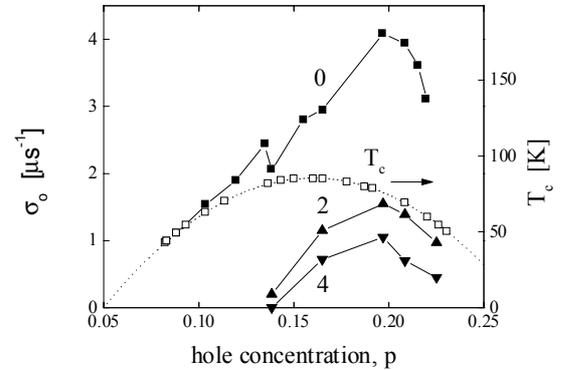

Fig. 4. The p-dependence of the μSR depolarisation rate ($\propto \lambda_{ab}^{-2}$) for $Y_{0.8}Ca_{0.2}Ba_2Cu_3O_{7-\delta}$ with 0, 2 and 4% Zn substitution on the $CuO_2$ planes.

## 6. Irreversibility field

6.1 Anisotropy dependence

HTS cuprates are distinguished in the wider class of superconductors by their anisotropy, defined as $\gamma=\lambda_c/\lambda_{ab}$, which may range from as small as $\gamma^2=6$ for Y-123 to $\gamma^2=50,000$ for Bi-2212 or Bi-2223 [19]. As a consequence of the large anisotropy the ab-plane pancake vortices readily decouple resulting in a marked reduction in pinning with consequent flux creep, reversible magnetisation and dissipative behaviour. This anisotropy is principally determined by the block-layer spacing, $d_b$, between the $CuO_2$ sheets. As already noted, to extract the structural effects on anisotropy it is essential to eliminate the strong effects of doping by investigating a broad range of cuprates each at the same doping state. Fig. 5 shows $H_{irr}(0.75T_c)$ plotted against $d_b$ for a representative range of cuprates all carefully controlled to optimal doping [3]. The correlation shows a strong exponential dependence of $H_{irr}(0.75T_c)$ on the $d_b$ spacing.

The data for Y-123 needs further explanation. Here the open data point in the figure represents the true block-layer spacing $d_b$=0.824nm. However, the chains in his compound are metallic and the SC order parameter extends onto the chains by proximity effect with a consequent doubling of the superfluid density [20]. As a result, the effective block-layer spacing is now the $CuO_2$-plane-to-CuO-chain spacing i.e. $d_b$ has halved and the solid data point for Y-123 is based on this effective $d_b$ value (0.412nm). It is the proximity-induced superconductivity on the metallic interlayer in this compound which gives it its remarkably high irreversibility field. Similarly high irreversibility fields are also observed in the case of Y-247. As a further check we investigated $H_{irr}$ for $Y_{0.8}Ca_{0.2}Ba_2Cu_3O_{7-\delta}$ at optimal doping with $\delta$=0.31. Here the oxygen ordering, and hence the chain metallicity, is destroyed and the effective $d_b$ spacing is now 0.824 nm. $H_{irr}$ is correspondingly reduced consistent with the exponential dependence upon $d_b$ as shown by the data point Ca-123. Finally the point Br-123 is for optimally-doped Y-123 with brominated non-metallic chains and is also consistent with the overall correlation amongst the cuprates [20].

It is remarkable that the only HTS materials currently used for long-length wire manufacture for magnets, cables etc are Bi-2212 and Bi-2223 which, as shown in Fig. 5, have the lowest irreversibility fields of all of the cuprates, a factor of 100 lower than for fully-oxygenated Y-123. Various models have been proposed for the irreversibility field including (i) the above-noted 3D-2D flux line transition involving decoupling of pancake vortices [21,22], (ii) a 3D melting transition involving a loss of static rigidity of the vortex lattice [23], (iii) bulk vortex depinning [15] and (iv) surface and geometric pinning effects [24]. The last two models are governed by the condensation energy which has much the same doping dependence and magnitude for all cuprates and do not reflect the dominant effect of anisotropy which gives the 100-fold varia-tion shown in Fig. 5. Blasius *et al.* [25] have investigated the vortex-matter phase diagram for Bi-2212 in the under-, optimal- and over-doped regions using µSR and magnetometry and concluded that the irreversibility field is governed by a 3D to 2D decoupling transition. Here $H_{irr} \propto \lambda_{ab}^{-1} \lambda_c^{-2} T^{-1}$ [22], and noting that $\lambda_c^{-2} \propto (\Delta'/\rho_c) \tanh[\Delta'/2kT]$ where $\Delta'=\Delta'(T/T_c)$ is the SC order parameter and $\rho_c(T)$ is the c-axis resistivity [26], then

$$H_{irr} \propto \lambda_{ab}(0)^{-1} [\Delta'(0)/T_c] \rho_c^{-1} f(t) \qquad (3)$$

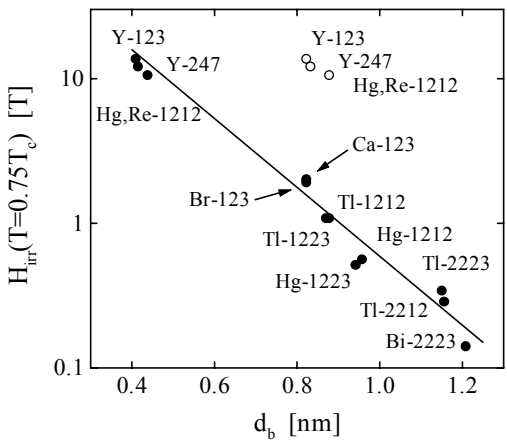

Fig. 5. The irreversibility field $H_{irr}(T=0.75T_c)$ for a range of HTS cuprates all at optimal-doping plotted as a function of the block-layer spacing $d_b$.

Where $f(t)$ is a universal function of $t=T/T_c$. We shall see this represents the experimental data for $H_{irr}$ very satisfactorily. Within a tunnelling model $\rho_c$ will vary as $\exp(d_b/d_o)$ where $d_o$ is a c-axis wave-function decay length. At a fixed doping state and fixed value of $t=0.75$ then $H_{irr} \propto \exp(-d_b/d_o)$ as indicated by Fig. 5. The value of $d_o$ is found to be 0.19nm.

6.2 Doping dependence

To explore the doping dependence of $H_{irr}$ we note that $[\Delta'(0)/T_c]$ in eqn. (1) will remain more or less constant, independent of doping. It is important to note the distinction here between the order parameter $\Delta'$ and the SC spectral gap $\Delta$ which, in the presence of the competing normal-state pseudogap are not the same. Based on the models of Bilbro and Macmillan [27] and Loram et al. [28] we may write $\Delta'^2 = \Delta^2 - E_g^2$. As shown in Fig. 1 $\Delta(0)$ remains more or less constant on the underdoped side and falls on the overdoped side with $T_c$. $2\Delta/kT_c$ remains constant on the overdoped side with the value 5.8 while, with the opening of the pseudogap, this ratio begins to rise rapidly as $T_c$ falls on the underdoped side. Thus the effect of the pseudogap is to reduce both $T_c$ and $\Delta'$ on the underdoped side with the result that $[\Delta'(0)/T_c]$ remains roughly constant even though $[\Delta(0)/T_c]$ diverges with underdoping. Eqn. (1) then simplifies to:

$$H_{irr} \propto \lambda_{ab}(0)^{-1} \rho_c^{-1} f(t) \qquad (4)$$

The p-dependence of $H_{irr}$ is thus governed by that of $\lambda_{ab}(0)$ and $\rho_c$. As noted in section 5 $\lambda_{ab}(0)^{-1}$ passes through a maximum at $p=0.19$ while $\rho_c$ has a strong doping dependence crossing over from semiconducting to metallic at $p=0.19$ due to the closure of the pseudogap [12]. Fig. 6 shows the doping dependence of $H_{irr}(0.7T_c)$ measured for single crystals of Bi-2212 (squares) and for polycrystalline samples of Bi-2223 (filled circles) [29]. Values of p were estimated from the $T_c$ values. The figure also shows the scaling field H* for the entire vortex matter phase diagram (including the irreversibility field) for single crystal Bi-2212 (diamonds) as reported by Blasius *et al.*, [25]. These values are compared with the model represented by equ. (2). For this we take the data of Fig. 4 for $\lambda_{ab}(0)^{-1}$ which, while for Y,Ca-123 appears to be generic for HTS cuprates. The c-axis resistivity data for Bi-2212 single crystals was reported by Watanabe *et al.* [30] and we fitted this to the form $C + \rho_{c,o}\exp(-p/p_o)$ in order to calculate $H_{irr}$. The correspondence between eqn. (2) and the $H_{irr}$ data is remarkably good even to the extent of duplicating the saturation that is apparent at large values of $p>0.2$. The overall p-dependence of $H_{irr}$ is thus seen to be dominated by the pseudogap both in the variation of $\lambda_{ab}(0)^{-1}$ and of $\rho_c$.

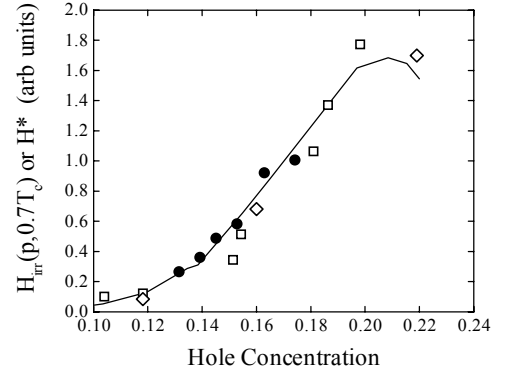

Fig. 6. The p-dependence of the irreversibility field $H_{irr}(T=0.7T_c)$ for Bi-2212 single crystals (squares), polycrystalline Bi-2223 (filled circles) and of the scaling field H* for Bi-2212 single crystals. The solid curve is calculated from the c-axis resistivity and $\lambda_{ab}^{-2}$ in Fig. 4 using eqn. (2).

7. Impurity scattering

It is well known that the HTS cuprates are very susceptible to the deleterious effects of impurities and defects on the $CuO_2$ planes, even for non-magnetic substituents such as Zn. The presence and impact of impurity and defect scattering in any HTS cuprate should not be judged in terms of its impact on $T_c$. It is a distinctive feature of a d-wave super-conductor that impurity scattering in the unitary limit results in a rapid suppression of the superfluid density, $\rho_s$, and a much slower suppression of $T_c$ [31]. A 10% reduction in $T_c$ results in a 27% reduction in $\rho_s$! More importantly still, the opening of the pseudogap is accompanied by a rapid increase in the suppression rate. This follows because within unitarity-limit scattering [32] the initial rate of suppression of $T_c$ is given by:

$$dT_c/dy = 58.5\, (\Delta_{oo}/T_{co})^{-1}\, \gamma^{-1} \qquad (5)$$

where y is the fraction of impurity substituted on the $CuO_2$ plane, $\gamma$ is the electronic specific heat coefficient in J/mol.K$^2$ and the second zero subscript refers to $\Delta_o$ and $T_c$ values when y=0. With the opening of the pseudogap, $\gamma$ near $T_c$ is progressively reduced so that $dT_c/dy$ is seen from eqn. (3) to be progressively increased i.e. both $T_c$ and the superfluid density become more rapidly suppressed due to impurities as the HTS cuprates become more underdoped, the pseudogap grows and $\gamma$ falls. Eqn. (3) describes the suppression in $T_c$ for Y:Ca-123, La-214 and Bi-2212 extremely well across the entire underdoped and overdoped phase curve with no adjustable parameters [32-34] simply using the experimentally-determined values of $\gamma$. In fact we used the average value of $\gamma(T)$ between 0 and $T_c$ taken directly from $S(T_c)/T_c$ which comes from the fact that $\gamma(T)=dS(T)/dT$, where $S(T)$ is the electronic entropy. The full Abrikosov-Gorkov theory result is shown in Fig. 7 for $Y_{0.8}Ca_{0.2}Ba_2Cu_3O_{7-\delta}$ with different values of $\delta$ spanning the entire phase curve [32]. Because $\gamma(T_c)$ is constant for p>0.19 the curves move down in parallel on the overdoped side but already at optimal doping (p=0.16) the curve $T_c(y)$ has steepened up due to the opening of the pseudogap. Further steepening occurs with progressive underdoping. The solid curves are the calculated Abrikosov-Gorkov curves using the experimentally-determined thermodynamic data and the pairbreaking strength obtained from eqn. (3).

One consequence of these considerations is that the phase curve $T_c(p)$ is not uniformly depressed with impurity substitution but is progressively displaced to higher doping so that it collapses around the pseudogap line $E_g(p)$ as shown in Fig. 8 for Bi-2212 [34]. The last point of superconductivity in critically Zn-substituted cuprates can be seen to occur at critical doping, p=0.19, which further underscores the fact that this is where superconductivity is most robust. This behaviour has been confirmed for Y-123 and La-214 [32,33].

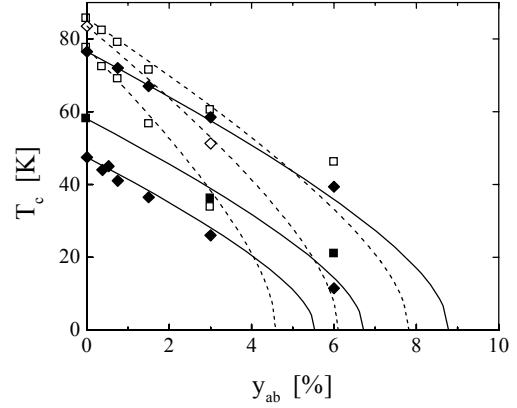

Fig. 7. The variation of $T_c$ with the concentration, $y_{ab}$, of Zn substituent on the $CuO_2$ planes for overdoped (solid curves and filled data points) and underdoped (dashed curves and open data points) for $Y_{0.8}Ca_{0.2}Ba_2Cu_3O_{7-\delta}$. The curves are the Abrikosov-Gorkov theory using the pairbreaking strength in eqn. (3) with measured values of the specific heat coefficient, $\gamma$, and no other adjustable parameters.

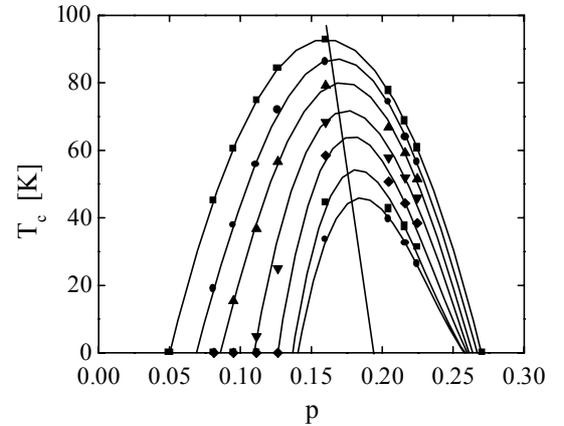

Fig. 8. The phase curves $T_c(p)$ for Bi-2212 with 0, 1, 2, 4, 6, 8 and 10% Co substitution. The sloping solid line is the pseudogap line, $E_g(p)$, which falls to zero at p=0.19.

## 8. Grain boundaries

Having seen the ubiquitous role of the pseudogap in determining condensation energy, superfluid density, critical currents and irreversibility field it is apparent that underdoping of the HTS cuprates should be avoided at all costs. It is not sufficient to quote a $T_c$ value around 90K in Y-123, one must establish that the material is fully oxygenated and hence overdoped close to the critical doping level. However, although the bulk may be critically doped, in polycrystalline samples grain boundaries tend to be underdoped due, for example, to incomplete oxygenation. This has been demonstrated by Babcock et al. [35] using TEM-EELS studies across individual grains boundaries. Local stresses at grain boundaries will tend to reduce the site energies for oxygen atoms in the CuO chains leading to oxygen deficiency, underdoping and the local presence of the pseudogap which will cause weakened intergranular critical currents. Moreover, the tendancy for impurities to accumulate in grain boundaries coupled with the local underdoping there will make these regions especially susceptible to impurity and defect scattering, strongly reducing the local superfluid density there. An obvious strategy for addressing grain-boundary underdoping due to reduced local site energies for oxygen is to anneal at rather low temperatures following oxygenation of the bulk at the usual temperatures of 400 to 450°C. This suggests a two-step anneal at 400 to 450C to oxygenate the bulk followed by 100 to 250°C to oxygenate the grain boundaries. Such temperatures would be too low for bulk oxygenation because of the low diffusion coefficient but not for grain boundaries where diffusion is much faster. The lower temperature limit is chosen on the basis that grain boundaries in Y-123 deoxygenate in an atmosphere of helium starting about 100 to 125°C [36].

## 9. Conclusions

It is apparent from the foregoing that the pseudogap has a profound influence upon critical currents and irreversibility fields in the HTS cuprates. The strong reduction of T=0 properties such as $U_o$, $J_c(T \to 0)$ and $\lambda_{ab}(0)^{-2}$ as well as the rapid onset of this reduction at p=0.19 make a phase-incoherent precursor pairing scenario rather unlikely. Moreover the onset of pseudogap effects occurs at the very point where the superfluid density is at its maximum suggesting that it is highly unlikely that the pseudogap effects are driven by the low superfluid density as has been suggested. The scaling of $J_c$ with $U_o$ and $H_{irr}(p)$ with $\lambda_{ab}(0)^{-2} \times \rho_c$ means that the ideal doping state for maximising these properties does not lie at optimum doping ($p \approx 0.16$) where $T_c$ maximises but at critical doping ($p \approx 0.19$) where the pseudogap disappears. This critical doping point may be found in any HTS cuprate by adjusting the thermoelectric power (TEP) to $S(290) = -2\mu V/K$ in contrast to optimal doping where $S(290) = +2\mu V/K$. The only exception so far found to this rule is La-214 which does not fit the observed TEP correlation with doping state [37]. The effect of the pseudogap is likely to be most deleterious at grain boundaries which tend to be underdoped and generally attract impurities and defects which rapidly suppress superfluid density in the presence of the pseudogap.

## Acknowledgements

Thanks are due to Dr. C. Bernhard and Dr. J. R. Cooper for ongoing discussion on the issues raised herein. This work was carried out under a James Cook Fellowship from the Royal Society of New Zealand and a Visiting Fellowship, Trinity College, Cambridge.